\documentclass[conference]{IEEEtran}
\IEEEoverridecommandlockouts
\usepackage{cite}
\usepackage{amsmath,amssymb,amsfonts}
\usepackage{algorithmic}
\usepackage{graphicx}
\usepackage{textcomp}
\def\BibTeX{{\rm B\kern-.05em{\sc i\kern-.025em b}\kern-.08em
    T\kern-.1667em\lower.7ex\hbox{E}\kern-.125emX}}

\usepackage{amsmath}
\usepackage{amsthm}
\usepackage{amssymb}
\usepackage{bm}
\usepackage{xspace}
\usepackage{xcolor}
\usepackage{graphicx}
\usepackage{url}
\usepackage{framed}
\usepackage{float}
\usepackage{rotating}
\usepackage{verbatim}
\usepackage{listings}
\usepackage{lscape}
\usepackage[ruled,vlined,linesnumbered]{algorithm2e}
\usepackage{caption}
\usepackage[font=footnotesize]{caption}
\usepackage{csquotes}

\usepackage{pgfplots}
\usepackage{pgf}
\usepackage{tikz}
\usetikzlibrary{arrows,shapes.misc,chains,scopes}
\usetikzlibrary{decorations.pathreplacing}

\usepackage{pgfplotstable}

\usepackage{colortbl}
\usetikzlibrary{matrix}
\usetikzlibrary{calc}
\usetikzlibrary{fit}
\usepackage{multicol}

\newcommand{\executeiffilenewer}[3]{%
	\ifnum\pdfstrcmp{\pdffilemoddate{#1}}%
	{\pdffilemoddate{#2}}>0%
	{\immediate\write18{#3}}\fi%
}

\graphicspath{{figures/}}

\newcommand{\bmm}{\begin{matrix}}
	\newcommand{\emm}{\end{matrix}}
\newcommand{\bpm}{\begin{pmatrix}}
	\newcommand{\epm}{\end{pmatrix}}

\newcommand{\bsbm}{\left[\begin{smallmatrix}}
	\newcommand{\esbm}{\end{smallmatrix}\right]}

\newcommand{\bbm}{\begin{bmatrix}}
	\newcommand{\ebm}{\end{bmatrix}}

\DeclareMathOperator*{\argmin}{argmin}
\DeclareMathOperator*{\argmax}{argmax}

\usepackage[utf8]{inputenc}
\usepackage[english]{babel}

\usepackage{booktabs}
\usepackage{multirow}
\usepackage{siunitx}
\usepackage{smartdiagram}

\theoremstyle{definition}

\begin{document}

\title{Construction and Decoding Algorithms for Polar Codes based on $2\times2$ Non-Binary Kernels}

\author{
	\IEEEauthorblockN{Peihong Yuan, Fabian Steiner}
	\IEEEauthorblockA{\textit{Institute for Communications Engineering} \\
	\textit{Technical University of Munich}\\
	Munich, Germany \\
	\{peihong.yuan, fabian.steiner\}@tum.de}
}

\maketitle

\begin{abstract}
	Polar codes based on $2\times2$ non-binary kernels are discussed in this work. The kernel over $\text{GF}(q)$ is selected by maximizing the polarization effect and using Monte-Carlo simulation. Belief propagation (BP) and successive cancellation (SC) based decoding algorithms are extended to non-binary codes. Additionally, a successive cancellation list (SCL) decoding with a pruned tree is proposed. Simulation results show that the proposed decoder performs very close to a conventional SCL decoder with significantly lower complexity. 
\end{abstract}


\section{Introduction}
Polar codes were proposed in~\cite{arikan2009channel,stolte2002rekursive} and they achieve the capacity of binary-input discrete memoryless channels asymptotically in the block length~\cite{arikan2009channel}. Due to their low complexity and excellent performance, polar codes have been adopted for the control channel in 5G enhanced mobile broadband (eMBB).

Successive cancellation (SC) was the first decoding algorithm for polar codes but has poor performance at short or moderate block length. Successive cancellation list (SCL) decoding~\cite{tal2015list} improves the finite length performance significantly at the cost of higher latency, power and storage consumption. Another approach to improve decoding is successive cancellation flip (SCF) decoding~\cite{afisiadis2014low}. SCF decoders try to identify and correct the first erroneous estimation during SC decoding by sequentially flipping the unreliable decisions. Belief propagation (BP) decoding algorithm for polar codes based on~\cite{forney2001codes} was proposed in~\cite{arikan2010polar}. In contrast to the SC-based approaches, BP decoding can provide a soft-output reliability information. Non-binary polar codes based on Reed-Solomon kernels are discussed in~\cite{mori2010non,trifonov2016polar}. The polarization theory for $q$-ary input channels are proposed in~\cite{csacsouglu2009polarization,chiu2014non}. In~\cite{chen2018new}, two-stage polarization for higher-order modulation is proposed.

In this work, we discuss polar codes over GF(q) based on the non-binary extension of the Ar{\i}kan kernel. In particular, we investigate non-binary polar codes over extension fields of GF($2$). The kernel is selected by Monte-Carlo simulation to maximize the polarization effect in the first stage. The SC-based and BP decoding algorithms are extended to non-binary symbols by using the message passing rule in the probability-domain proposed in~\cite{davey1998low}. The fast Fourier transform (FFT) based check node (CN) operation~\cite{barnault2003fast} reduces the CN complexity from $\mathcal{O}(q^2)$ to $\mathcal{O}(q\log_2q)$. Additionally, an SCL algorithm with a pruned tree is proposed. Simulation results show that the proposed decoder performs very close to the conventional SCL decoder with significantly lower average computational complexity.

This work is organized as follows. In Sec.~\ref{sec:nbpolar}, non-binary SC decoding and the Monte-Carlo based kernel selection are introduced. In Sec.~\ref{sec:dec}, we extend BP, SCF and SCL decoding to the non-binary case and discuss the simulation results. SCL with a pruned tree is proposed in Sec.~\ref{subsec:prun}. We conclude in Sec.~\ref{sec:con}.

\section{Polar Codes over $\text{GF}(q)$}\label{sec:nbpolar}

\subsection{Notation}
Uppercase letters denote random variables (RVs) while the corresponding lowercase letters are their realizations. The notation $c^n$ denote a vector of length $n$. $\boldsymbol{P}_X$ denotes the probability mass function (PMF) of a discrete RV $X$. We consider GFs of order $q=2^r,r>1$, i.e., extension fields of GF($2$). Binary and decimal representations are used to describe elements over GFs, i.e., GF($q$) has $q$ elements and the elements can be represented by binary $r$-tuples or integers between $0$ and $q-1$. The primitive polynomials \cite{berlekamp1968algebraic} of extension fields are represented by decimals.

For the codes over GF($q$), $n_c$ denotes the code length in symbols, $n_{c,\text{bin}}=n_cr$ denotes the code length in bits, $k_{c,\text{bin}}$ is the code dimension in bits. 
\subsection{System Model}
A non-binary polar code over GF($q$) of length $n_c$ and dimension $k_{c,\text{bin}}$ is defined by the $q$-ary polar transform $\mathbb{F}^{\otimes \log_2n_c}$ and $n_{c,\text{bin}} - k_{c,\text{bin}}$ frozen (bit) positions, where $\mathbb{F}$ denotes the extended Ar\i kan kernel
\begin{equation}
	\mathbb{F}=\begin{bmatrix}
	1 & 0\\
	\alpha & \beta
	\end{bmatrix},\qquad \alpha,\beta\in\text{GF}(q)
\end{equation}
and $(\cdot)^\otimes$ denotes the Kronecker power. Polar encoding can be represented by $c^{n_c}=u^{n_c}\mathbb{F}^{\otimes \log_2{n_c}}.$
The vectors $u^{n_c}$, $c^{n_c}$ and all (addition, multiplication) operations are defined over GF($q$). The vector $c^{n_c}$ denotes the code symbols. The vector $u^{n_c}$ can be represented by $n_{c,\text{bin}}$ bits and includes information bits and frozen bits. For a symbol $u_i\in\text{GF}(q)$, the first $t_i$ bits can be selected as frozen, where $t_i=0,\dots,r$. The choice of symbol $u_i$ is then restricted to $0,\dots,{2^{r-t_i}}-1$ and the symbol carries $r-t_i$ bits of information (assuming that the left bit is most significant for the bit-to-symbol conversion). We define the set for all possibilities of symbol $u_i$ as $\mathcal{S}(u_i)=\left\{0,\dots,{2^{r-t_i}}-1\right\}.$
If $t_i=0$, then $|\mathcal{S}(u_i)|=q$. If $t_i=r$, then $\mathcal{S}(u_i)=\left\{0\right\}$. We have $\sum_{i=1}^{n_c}\log_2 |\mathcal{S}(u_i)|=k_{c,\text{bin}}.$ The code rate is given by $k_{c,\text{bin}}/n_{c,\text{bin}}$.
	
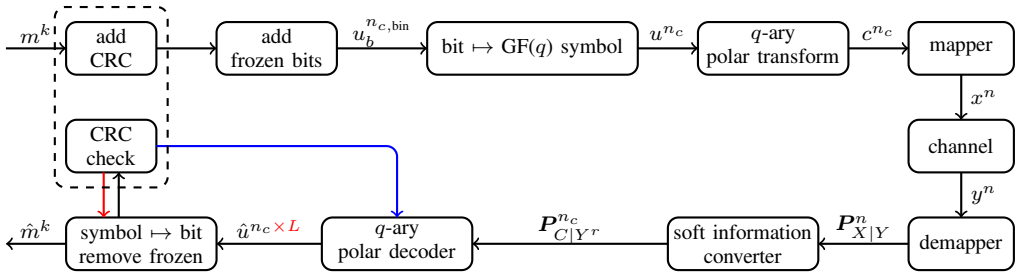
\begin{figure*}
		\centering
		\begin{tikzpicture}
		\footnotesize
		\node at (-0.4,0.55) {$m^k$};
		\draw[->,thick] (-0.8,0.35) to (0,0.35);
		\draw[rounded corners,thick]   (0,0) rectangle (1.2,0.7);
		\node at (0.6,0.35)[align=center]  {add \\ CRC};
		\draw[->,thick] (1.2,0.35) to (2.0,0.35);
		\draw[rounded corners,thick]   (2.0,0) rectangle (3.6,0.7);
		\node at (2.8,0.35)[align=center] {add \\ frozen bits};
		\draw[->,thick] (3.6,0.35) to (4.8,0.35);
		\node at (4.2,0.55) {$u_b^{n_{c,\text{bin}}}$};
		\draw[rounded corners,thick]   (4.8,0) rectangle (7.6,0.7);
		\node at (6.2,0.35)[align=center] {bit $\mapsto$ GF($q$) symbol};
		\draw[->,thick] (7.6,0.35) to (8.4,0.35);
		\node at (8,0.55) {$u^{n_c}$};
		\draw[rounded corners,thick]   (8.4,0) rectangle (10.4,0.7);
		\node at (9.4,0.35)[align=center] {$q$-ary\\ polar transform};
		\draw[->,thick] (10.4,0.35) to (11.2,0.35);
		\node at (10.8,0.55) {$c^{n_c}$};
		\draw[rounded corners,thick]   (11.2,0) rectangle (12.6,0.7);
		\node at (11.9,0.35)[align=center] {mapper};
		\draw[->,thick] (11.9,0) to (11.9,-0.6);
		\node at (12.2,-0.3) {$x^{n}$};
		\draw[rounded corners,thick]   (11.2,-1.3) rectangle (12.6,-0.6);
		\node at (11.9,-0.95)[align=center] {channel};
		\draw[->,thick] (11.9,-1.3) to (11.9,-1.9);
		\node at (12.2,-1.6) {$y^{n}$};
		\draw[rounded corners,thick]   (11.2,-2.6) rectangle (12.6,-1.9);
		\node at (11.9,-2.25)[align=center] {demapper};
		\draw[->,thick] (11.2,-2.25) to (10,-2.25);
		\node at (10.6,-2.05) {$\boldsymbol{P}_{X|Y}^{n}$};
		\draw[rounded corners,thick]   (8,-2.6) rectangle (10,-1.9);
		\node at (9,-2.25)[align=center] {soft information\\converter};
		\draw[->,thick] (8,-2.25) to (5.4,-2.25);
		\node at (6.7,-2.05) {$\boldsymbol{P}_{C|Y^{r}}^{n_c}$};
		\draw[rounded corners,thick]   (3.4,-2.6) rectangle (5.4,-1.9);
		\node at (4.4,-2.25)[align=center] {$q$-ary\\ polar decoder};
		\draw[->,thick] (3.4,-2.25) to (2,-2.25);
		\node at (2.7,-2.05) {$\hat{u}^{n_c \textcolor{red}{\times L} }$};
		\draw[rounded corners,thick]   (0,-2.6) rectangle (2,-1.9);
		\node at (1,-2.25)[align=center] {symbol $\mapsto$ bit\\remove frozen};
		\draw[->,thick] (0,-2.25) to (-0.8,-2.25);
		\node at (-0.4,-2.05) {$\hat{m}^k$};
		\draw[rounded corners,thick]   (0,-1.3) rectangle (1.2,-0.6);
		\node at (0.6,-0.95)[align=center] {CRC\\check};
		\draw[->,thick,red] (0.5,-1.3) to (0.5,-1.9);
		\draw[<-,thick] (0.7,-1.3) to (0.7,-1.9);
		\draw[thick,rounded corners,blue,->] (1.2,-0.95) -- (4.4,-0.95) -- (4.4,-1.9);
		\draw[rounded corners,thick,dashed]   (-0.15,-1.5) rectangle (1.35,0.9);
		\end{tikzpicture}
		\caption{The transmission system with $q$-ary polar codes over binary input AWGN channel. The red part is only for SCL decoding and the blue part only for SCF decoding. $\boldsymbol{P}_{X|Y}$ and $\boldsymbol{P}_{C|Y^{r}}$ are PMFs and consist of vectors of $2$ and $q$ probabilities.}
		\label{fig:system}
	\end{figure*}

The transmission system with $q$-ary polar codes is shown in Fig.~\ref{fig:system}. The vector $u_b^{n_{c,\text{bin}}}$ includes $k$ information bits $m^k$, $\ell_\text{CRC}$ cyclic redundancy check (CRC) bits and $n_{c,\text{bin}}-k-\ell_\text{CRC}$ frozen bits (preset to zero). We have  $k_{c,\text{bin}}=k+\ell_\text{CRC}$ in this case. $u_b^{n_{c,\text{bin}}}$ is then converted into the symbol vector $u^{n_c}$. For a binary input channel, the code symbol $c_i$ is mapped into $r$ channel symbols: 
\begin{align}
c_i &\mapsto x_{i,1},\dots,x_{i,r}=x_{i}^{r},~i=1,\dots,n_c\\
c^{n_c} &\mapsto \left(x^{r}\right)^{n_c} =x^{n}
\end{align}
where $n=n_cr=n_{c,\text{bin}}$ denotes the number of channel uses. 

The discrete time memoryless additive white Gaussian noise (AWGN) channel with binary input is described by $Y_j = X_j + \sigma Z_j$, $j=1,\dots,n$, where $X\in\mathcal{X}=\left\{-\Delta,+\Delta\right\}$ and $Z^{n}$ is a string of $n$ independent and identically distributed zero mean Gaussian RVs with variance one. The signal-to-noise ratio $E_s/N_0$ is given by $\mathbb{E}(X^2)/\sigma^2=\Delta^2/\sigma^2$, where $\mathbb{E}(\cdot)$ denotes expectation. At the receiver, the demapper computes the a posteriori probabilities (APPs) of $X$ via
\begin{equation}
P_{X|Y}(x_j|y_j)=P_{X}(x_j)\frac{p_{Y|X}(y_j|x_j) }{ \sum_{a\in\mathcal{X}}P_{X}(a) p_{Y|X}(y_j|a)}
\end{equation}
where $p_{Y|X}(y_j|x_j)=\frac{1}{\sqrt{2\pi\sigma^2}}\exp{\frac{(y_j-x_j)^2}{2\sigma^2}}.$
Let $\boldsymbol{P}_{C|Y^r,i}$ denote the PMF of code symbol $C_i$ given the relevant channel outputs: 
\begin{equation}
\boldsymbol{P}_{C|Y^{r},i}=\left[P_{C|Y^{r}}(0|y_{i}^{r}),  \dots,P_{C|Y^{r}}(q-1|y_{i}^{r}) \right]
\end{equation}
where $P_{C|Y^{r}}(c_i|y_{i}^{r}) = \prod_{v=1}^{r}P_{X|Y}(x_{i,v}|y_{i,v}).$ The PMFs of $n_c$ code symbols $\boldsymbol{P}_{C|Y^{r}}^{n_c}$ are delivered to the decoder. The decoder outputs the estimated information bits $\hat{m}^{k}$ with the help of the CRC. The transmission rate is $k/n$.

Due to limitations of space, we only consider binary input AWGN channel in this paper.
\subsection{Message Passing on Non-Binary Graphs}\label{sec:mp}
%
%
%
Assume that the BP decoder operates in the probability domain. Then, each message $A$ can be described as an RV with a PMF, i.e.,
$
\boldsymbol{P}_A = \left[P_A(0),P_A(1)\dots,P_A(q-1) \right].
$
We have three basic probability domain operations:
\begin{itemize}
	\item Multiplication and addition:\\
	Suppose $B=A\alpha$ with $\alpha \in\text{GF(q)}$. We then have for the PMF $P_A(\mu) = P_B(\mu\alpha)$, $\mu\in\text{GF}(q)$ and can find a $q\times q$ permutation matrix $\Pi_{\cdot\alpha}$ such that
	\begin{align}
	\boldsymbol{P}_{B} = \boldsymbol{P}_{A} \boldsymbol{\Pi}_{\cdot\alpha},~\boldsymbol{P}_{A} = \boldsymbol{P}_{B} \boldsymbol{\Pi}_{\cdot\alpha}^{-1}.
	\end{align}
	We can also find a permutation matrix $\boldsymbol{\Pi}_{+\alpha}$ for addition
\begin{equation}
\boldsymbol{P}_{A+\alpha} = \boldsymbol{P}_{A} \boldsymbol{\Pi}_{+\alpha}.
\end{equation}
Note that $\boldsymbol{\Pi}_{\cdot\alpha}$ and $\boldsymbol{\Pi}_{+\alpha}$ depend on the primitive polynomial. 
	\item Check node (CN) update:\\
	The CN node computes the PMF $\boldsymbol{P}_{A+B}$ from $\boldsymbol{P}_{A}$ and $\boldsymbol{P}_{B}$.
	\begin{equation}
	P_{A+B}(\mu)=\sum_{\mu_1,\mu_2:\mu_1+\mu_2=\mu}P_{A}(\mu_1)P_{B}(\mu_2).
	\end{equation}
	We have $\boldsymbol{P}_{A+B}=\boldsymbol{P}_{A}\circledast\boldsymbol{P}_{B}$, where $\circledast$ denotes the cyclic discrete convolution and the complexity is given by $\mathcal{O}(q^2)$~\cite{davey1998low}. By applying the fast Hadamard transform (FHT), the discrete convolution is translated to element-wise multiplication, which reduces the complexity of the CN update to $\mathcal{O}(q\log_2q)$. We have
	\begin{equation}
	\boldsymbol{P}_{A+B}=\mathcal{H}\left( \mathcal{H}(\boldsymbol{P}_{A})\odot \mathcal{H}(\boldsymbol{P}_{B}) \right)
	\end{equation}
	where $\mathcal{H}(\cdot)$ denotes the FHT operation and $\odot$ denotes the element-wise multiplication. Note that the FHT is self inverse, i.e., $\mathcal{H(\cdot)}=\mathcal{H}^{-1}(\cdot)$. 
	\item Variable node (VN) update:\\
	The VN node computes the PMF of $A$ from two different observed PMFs $\boldsymbol{P}_{A,1}$ and $\boldsymbol{P}_{A,2}$. We have 
	$P_{A}(\mu)= P_{A,1}(\mu)P_{A,2}(\mu)$, $\mu\in\text{GF}(q)$
	and
		\begin{equation}
		\boldsymbol{P}_{A} = \boldsymbol{P}_{A,1}\odot\boldsymbol{P}_{A,2}.
		\end{equation}
\end{itemize}
The elements in the output message from the CNs and VNs must be normalized.

\subsection{SC Decoding}
The non-binary SC decoder follows mainly the implementation in~\cite[Algorithm 2-4]{tal2015list}. Fig.~\ref{fig:dec} shows the extended recursive message update rules. We have
\begin{align}
\boldsymbol{P}_1^\prime&=a_1\mathcal{H}\Big( \mathcal{H}(\boldsymbol{P}_1)\odot \mathcal{H}(\boldsymbol{P}_2\boldsymbol{\Pi}_{\cdot\beta}^{-1}\boldsymbol{\Pi}_{\cdot\alpha}) \Big) \nonumber\\
&=a_1\mathcal{H}\Big( \mathcal{H}(\boldsymbol{P}_1)\odot \mathcal{H}(\boldsymbol{P}_2\boldsymbol{\Pi}_{\cdot\frac{\alpha}{\beta}}) \Big)\label{eq:dec_c}\\
\boldsymbol{P}_2^\prime&=a_2\Big( \boldsymbol{P}_1\boldsymbol{\Pi}_{+\hat{u}_1}\boldsymbol{\Pi}^{-1}_{\cdot\alpha} \Big) \odot \Big( \boldsymbol{P}_2\boldsymbol{\Pi}^{-1}_{\cdot\beta} \Big)\label{eq:dec_v}
\end{align}
where the scalar $a_1$ and $a_2$ are the normalization factors that ensure that the probabilities in $\boldsymbol{P}_1^\prime$ and $\boldsymbol{P}_2^\prime$ sum up to 1. At the decoding phase $i$ ($1\leq i\leq n_c$), we get a conditional PMF of $U_i$ by recursively updating the messages
\begin{align}\label{eq:sc}
\boldsymbol{P}_{U_i|Y^nU^{i-1}}=\left[\dots,P_{U_i|Y^nU^{i-1}}\left(\mu| y^n\hat{u}^{i-1}\right),\dots\right],~\mu\in\text{GF}(q)
\end{align}
The hard decision of $u_i$ is given by
\begin{equation}
\hat{u}_i = \argmax_{\mu\in\mathcal{S}(u_i)} P_{U_i|Y^nU^{i-1}}\left(\mu| y^n\hat{u}^{i-1}\right).
\end{equation}
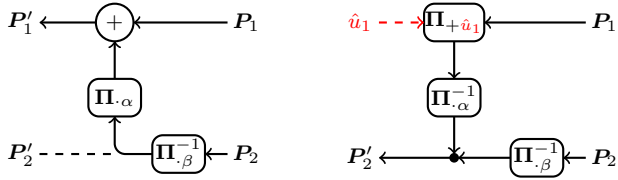
\begin{figure}
	\centering
	\begin{multicols}{2}
	\begin{tikzpicture}
	\footnotesize
	\draw [thick] (0,0) circle [radius=0.25];
	\node at (0,0) {$+$};
	\draw[<-,thick] (-1,0) to (-0.25,0);
	\draw[<-,thick] (0.25,0) to (1.5,0);
	\draw[->,thick] (0,-0.75) to (0,-0.25);
	\draw[rounded corners,thick]   (-0.35,-1.25) rectangle (0.35,-0.75);
	\node at (0,-1) {$\boldsymbol{\Pi}_{\cdot\alpha}$};
	\draw[thick,rounded corners,->] (0.5,-1.75) -- (0,-1.75) -- (0,-1.25);
	\draw[dashed,thick] (-1,-1.75) to (0.0,-1.75);
	\draw[<-,thick] (1.2,-1.75) to (1.5,-1.75);
	\draw[rounded corners,thick]   (0.5,-2) rectangle (1.2,-1.5);
	\node at (0.85,-1.75) {$\boldsymbol{\Pi}^{-1}_{\cdot\beta}$};
	\node at (-1.25,0) {$\boldsymbol{P}^\prime_1$};
	\node at (1.75,0) {$\boldsymbol{P}_1$};
	\node at (-1.25,-1.75) {$\boldsymbol{P}^\prime_2$};
	\node at (1.75,-1.75) {$\boldsymbol{P}_2$};
	\end{tikzpicture}
	
	\begin{tikzpicture}
	\footnotesize
	\draw[rounded corners,thick]   (-0.4,-0.25) rectangle (0.4,0.25);
	\node at (0,0) {$\boldsymbol{\Pi}_{+\textcolor{red}{\hat{u}_1}}$};
	\draw[->,thick,red,dashed] (-1,0) to (-0.4,0);
	\draw[<-,thick] (0.4,0) to (1.75,0);
	\draw[<-,thick] (0,-0.75) to (0,-0.25);
	\draw[rounded corners,thick]   (-0.35,-1.25) rectangle (0.35,-0.75);
	\node at (0,-1) {$\boldsymbol{\Pi}^{-1}_{\cdot\alpha}$};
	\draw[->,thick] (0,-1.25) to (0,-1.75);
	\draw [thick,fill=black] (0,-1.8) circle [radius=0.05];
	\draw[<-,thick] (-1,-1.8) to (-0.05,-1.8);
	\draw[<-,thick] (1.45,-1.8) to (1.75,-1.8);
	\draw[rounded corners,thick]   (0.75,-2.05) rectangle (1.45,-1.55);
	\node at (1.1,-1.8) {$\boldsymbol{\Pi}^{-1}_{\cdot\beta}$};
	\draw[<-,thick] (0.05,-1.8) to (0.75,-1.8);
	\node at (-1.25,0) {$\textcolor{red}{\hat{u}_1}$};
	\node at (2,0) {$\boldsymbol{P}_1$};
	\node at (-1.25,-1.8) {$\boldsymbol{P}^\prime_2$};
	\node at (2,-1.8) {$\boldsymbol{P}_2$};
	\end{tikzpicture}
	\end{multicols}
	\caption{Message update rules for $q$-ary SC decoder.}
	\label{fig:dec}
\end{figure}
\subsection{Kernel Selection}
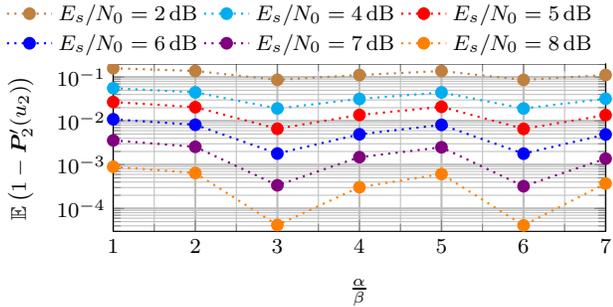
\begin{figure}
	\centering
	\begin{tikzpicture}[scale=1]
	\footnotesize
	\begin{semilogyaxis}[
	legend style={at={(1,1)},anchor=south east,draw=none},
	legend columns=3, 
	width=3.2in,
	height=1.5in,
	ymin=0.00003,
	ymax=0.2,
	minor x tick num=1,
	minor y tick num=4,
	major grid style={thick},
	grid=both,
	xmin = 1,
	xmax = 7,
	xtick={1,2,3,4,5,6,7},
	xlabel = $\frac{\alpha}{\beta}$,
	ylabel = $\mathbb{E}\left(1-\boldsymbol{P}_2^\prime(u_2)\right)$ ,
	]		
		\addplot[brown, mark = *,thick,mark options={solid},dotted]
		table[x=snr,y=fer]{snr fer
			1 0.15891
			2 0.13758
			3 0.086417
			4 0.11127
			5 0.13735
			6 0.086061
			7 0.11154
			};\addlegendentry{$E_s/N_0=\SI{2}{dB}$}
		
		\addplot[cyan, mark = *,thick,mark options={solid},dotted]
		table[x=snr,y=fer]{snr fer
			1 0.055799
			2 0.044716
			3 0.01917
			4 0.031669
			5 0.044578
			6 0.01918
			7 0.031969
			};\addlegendentry{$E_s/N_0= \SI{4}{dB}$}
			
				\addplot[red, mark = *,thick,mark options={solid},dotted]
				table[x=snr,y=fer]{snr fer
					1 0.026978
					2 0.020524
					3 0.0065985
					4 0.01364
					5 0.020908
					6 0.006554
					7 0.013525
				};\addlegendentry{$E_s/N_0=\SI{5}{dB}$\quad}
		
		\addplot[blue, mark = *,thick,mark options={solid},dotted]
		table[x=snr,y=fer]{snr fer
			1 0.010865
			2 0.0081376
			3 0.001793
			4 0.0049267
			5 0.0081024
			6 0.0017836
			7 0.0048821
		};\addlegendentry{$E_s/N_0=\SI{6}{dB}$}
		
		\addplot[violet, mark = *,thick,mark options={solid},dotted]
		table[x=snr,y=fer]{snr fer
			1 0.003583
			2 0.0025569
			3 0.00033801
			4 0.0014807
			5 0.0024915
			6 0.00032161
			7 0.0013711
		};\addlegendentry{$E_s/N_0=\SI{7}{dB}$}
		
		\addplot[orange, mark = *,thick,mark options={solid},dotted]
		table[x=snr,y=fer]{snr fer
			1 0.00089431
			2 0.00064435
			3 4.1891e-05
			4 0.00030564
			5 0.00061048
			6 4.1129e-05
			7 0.00037214
			
		};\addlegendentry{$E_s/N_0=\SI{8}{dB}$}
		
 	\end{semilogyaxis}
	\end{tikzpicture}
	\caption{Reliability of \enquote{single-level} polarization over GF($8$), $\text{primitive polynomial}=11$}
	\label{fig:kernel_gf8}
\end{figure}

From (\ref{eq:dec_c}) and (\ref{eq:dec_v}), we observe that the $\boldsymbol{P}_1^\prime(u_1)$ and $\boldsymbol{P}_2^\prime(u_2)$ depend only on the ratio $\alpha/\beta$. We now use a Monte-Carlo approach to choose the best ratio $\alpha/\beta$:
\begin{itemize}
	\item[1.] Set $u_1=0$ and select $u_2\in\text{GF}(q)$ randomly.
	\item[2.] Encode ($n_c=2$): $c_1=u_1+u_2\alpha$ and $c_2=u_2\beta$.
	\item[3.] Map $c_1$ and $c_2$ to the $n$ channel symbols ($n=2r$).
	\item[4.] Add noise $\sigma Z^{n}$ and compute $\boldsymbol{P}_2^\prime(u_2)$.
\end{itemize}
Note that $\boldsymbol{P}_2^\prime(u_2)$ is now a RV depending on $\sigma Z^{n}$. Monte-Carlo simulation is used to find the optimal $\alpha/\beta$ which maximizes the \enquote{single-level} polarization effect, i.e., we choose
\begin{equation}
\frac{\alpha}{\beta} = \argmin_{\frac{\alpha}{\beta}\in\text{GF}(q)} \mathbb{E}\left(1-\boldsymbol{P}_2^\prime(u_2)\right).
\end{equation}
An example for GF($8$) with binary input AWGN (biAWGN) is shown in Fig.~\ref{fig:kernel_gf8}. We observe that the ratios $\alpha/\beta=3$ and $\alpha/\beta=6$ provide the strongest \enquote{single-level} polarization for any channel qualities. We summarize good kernels for GFs for other field orders in Table~\ref{tab:kernels}.
\begin{table}
	\footnotesize
	\centering
	\caption{Good kernels for GF($q$), $q=\left\{4,8,16,32,64,128,256\right\}$}
	\label{tab:kernels}
	{\renewcommand{\arraystretch}{1.5}
		\begin{tabular}{|c|c|c|}				
			\hline
			$q$& Primitive Polynomial & $\alpha/\beta$ \\
			\hline
			$4$ & $7$&  $2,3$\\
			\hline
			$8$ & $11$&  $3,6$\\
			\hline
			$16$ & $19$&  $6,7$\\
			\hline
			$32$ & $37$&  $13,15,21,26$\\
			\hline
			$64$ & $67$&  $38,50$\\
			\hline
			$128$ & $137$ &  $57,105$\\
			\hline
			$256$ & $285$&  $23,29,102,131,133,81,145,212$\\
			\hline
		\end{tabular}
	}
\end{table}	

\section{Improved Decoders for Non-Binary Polar Codes}\label{sec:dec}
\subsection{BP Decoding}
BP decoding of non-binary polar codes is a message passing algorithm on the encoding graph. We use the flooding schedule in~\cite[Sec. II-B]{cammerer2017combining} with the $q$-ary message passing rules described in Sec.~\ref{sec:mp}. An early stopping criteria~\cite{yuan2014early} is used.

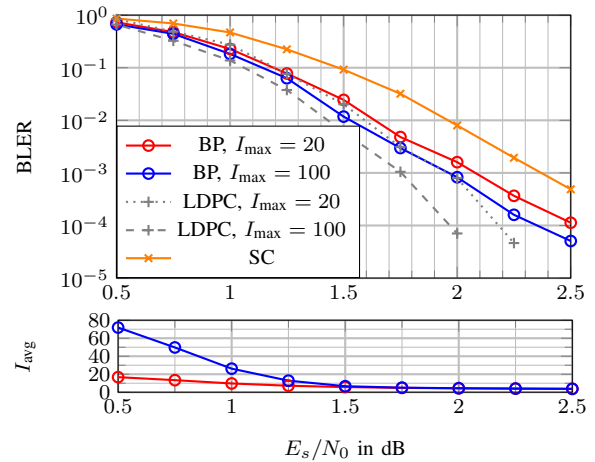
\begin{figure}
	\centering
	\begin{tikzpicture}[scale=1]
	\footnotesize
	\begin{semilogyaxis}[
	legend style={at={(0,0)},anchor=south west},
	ymin=0.00001,
	ymax=1,
	width=3in,
	height=2in,
	minor x tick num=4,
	minor y tick num=5,
	major grid style={thick},
	ytick={1,0.1,0.01,0.001,0.0001,0.00001,0.000001},
	grid=both,
	xmin = 0.5,
	xmax = 2.5,
	ylabel = BLER,
	]
	\addplot[red, mark = o,thick]
	table[x=snr,y=fer]{snr fer
		0 0.96774
		0.25 0.96774
		0.5 0.73171
		0.75 0.47619
		1 0.22727
		1.25 0.07772
		1.5 0.02447
		1.75 0.0048123
		2 0.001601
		2.25 0.00036688
		2.5 0.00011297
	};\addlegendentry{BP, $I_\text{max}=20$}
	
	\addplot[blue, mark = o,thick]
	table[x=snr,y=fer]{snr fer
		0 0.9375
		0.25 0.96774
		0.5 0.66667
		0.75 0.44118
		1 0.18293
		1.25 0.063158
		1.5 0.011765
		1.75 0.00297
		2 0.00082711
		2.25 0.0001597
		2.5 5.0735e-05
	};\addlegendentry{BP, $I_\text{max}=100$}
	
	\addplot[mark options={solid}, mark = +,gray,dotted,thick]
	table[x=snr,y=fer]{snr fer ser frames
		0.250000 9.138e-01 4.464e-01 58 2.060e+01
		0.500000 8.413e-01 3.502e-01 63 1.984e+01
		0.750000 4.860e-01 1.849e-01 107 1.665e+01
		1.000000 2.751e-01 7.486e-02 189 1.340e+01
		1.250000 7.657e-02 2.043e-02 653 9.741e+00
		1.500000 2.013e-02 4.431e-03 2484 7.473e+00
		1.750000 3.106e-03 5.377e-04 16098 5.843e+00
		2.000000 7.943e-04 1.036e-04 62951 4.845e+00
		2.250000 4.640e-05 4.560e-06 1077640 4.139e+00
	};\addlegendentry{LDPC, $I_\text{max}=20$}
	
	\addplot[mark options={solid}, mark = +,gray,dashed,thick]
	table[x=snr,y=fer]{snr fer ser frames
		0.250000 8.413e-01 4.235e-01 63 8.949e+01
		0.500000 6.625e-01 3.397e-01 80 7.369e+01
		0.750000 3.250e-01 1.355e-01 160 4.476e+01
		1.000000 1.375e-01 5.214e-02 371 2.538e+01
		1.250000 3.776e-02 1.441e-02 1324 1.347e+01
		1.500000 5.993e-03 1.800e-03 8510 7.973e+00
		1.750000 1.050e-03 2.804e-04 47636 5.973e+00
		2.000000 7.057e-05 1.595e-05 708566 4.846e+00
	};\addlegendentry{LDPC, $I_\text{max}=100$}
	
	\addplot[orange, mark = x,thick]
	table[x=snr,y=fer]{snr fer
		0 1
		0.25 0.97059
		0.5 0.86047
		0.75 0.69492
		1 0.46847
		1.25 0.22394
		1.5 0.09212
		1.75 0.031973
		2 0.0080037
		2.25 0.001941
		2.5 0.000488
		2.75 0.0001
	};\addlegendentry{SC}
	
	\end{semilogyaxis}
	\end{tikzpicture}
	\begin{tikzpicture}[scale=1]
	\footnotesize
	\begin{axis}[
	width=3in,
	height=1in,
	legend style={at={(1,1)},anchor=north east},
	legend columns=1, 
	ymin=0,
	ymax=80,
	minor x tick num=1,
	minor y tick num=1,
	major grid style={thick},
	grid=both,
	xmin = 0.5,
	xmax = 2.5,
	xlabel = $E_s/N_0\ \text{in dB}$,
	ylabel = $I_\text{avg}$,
	]
	\addplot[red, mark = o,thick]
	table[x=snr,y=fer]{snr fer
		0 19.968
		0.25 19.613
		0.5 16.756
		0.75 13.413
		1 9.7424
		1.25 7.2254
		1.5 5.7325
		1.75 4.8266
		2 4.3633
		2.25 4.0323
		2.5 3.7684
	};
	\addplot[blue, mark = o,thick]
	table[x=snr,y=fer]{snr fer
		0 96.594
		0.25 97.032
		0.5 71.867
		0.75 49.75
		1 26.195
		1.25 12.863
		1.5 6.6843
		1.75 5.0723
		2 4.4199
		2.25 4.0408
		2.5 3.7689
	};
	\end{axis}
	\end{tikzpicture}
	\caption{BP decoding performance of polar codes over GF($256$), $n_c=128$, $k=512$, $\text{primitive polynomial}=285$, $\alpha=29$, $\beta=1$, frozen position designed by Monte-Carlo method for SC decoding at $\SI{2.5}{dB}$}
	\label{fig:bp}
\end{figure}

The block error rate (BLER) and average number of iterations ($I_\text{avg}$) of BP decoding with $I_\text{max}=20$ and $100$ is shown in Fig.~\ref{fig:bp}. We observe that the BP decoding ($I_\text{max}=100$) outperforms SC decoding by $\SI{0.4}{dB}$ with low average complexity/latency ($I_\text{avg}\leq 5$) in the high SNR regime ($\geq\SI{1.75}{dB}$).

As a reference, we also provide simulation results for ultra sparse non-binary LDPC codes (with regular variable and check node degrees of two and four, respectively)~\cite{poulliat_design_2008} over GF($256$) with the same parameters. We observe that the non-binary LDPC code outperforms the polar code by $\SI{0.5}{dB}$ with BP decoding ($I_\text{max}=100$).

\subsection{SCF Decoding}
Due to the serial nature of SC decoding, an erroneous bit decision can be caused by the channel noise or previous erroneous bit estimates. The main idea of SCF decoding~\cite{afisiadis2014low} is trying to correct the first erroneous bit decision by sequentially flipping the unreliable decisions. The authors in~\cite{afisiadis2014low} use an oracle-assisted SC decoder (SCO-1) to describe the potential benefits of correcting the first error. SCO-1 is an SC decoder that can identify and correct the first bit error. In the non-binary case, an SCF decoder tries to correct the first erroneous decision for the  $q$-ary symbols. Fig.~\ref{fig:scf_errors} shows the histogram of the number of errors caused by the channel noise. We observe that $98.1\%$ of the block errors at $\SI{2.25}{dB}$ are corrected by the SCO-1 decoder. 

Consider $q$-ary polar codes with rate $(k+\ell_\text{CRC})/n_cr=(k+\ell_\text{CRC})/n_{c,\text{bin}}$. We use an $\ell_\text{CRC}$ bits CRC outer code to check whether the output is a valid codeword or not. The SCF decoder starts by performing SC decoding to generate the first estimation $\hat{u}^{n_c}$ and stores the soft information in an $n_c\times\left(q-1\right)$ matrix $\boldsymbol{\varLambda}$:
\begin{align}
\boldsymbol{\varLambda}_{i,\mu}=\left\{\begin{aligned}
&\frac{P_{U_i|Y^nU^{i-1}}\left(\mu| y^n\hat{u}^{i-1}\right)}{P_{U_i|Y^nU^{i-1}}\left(\hat{u}_i| y^n\hat{u}^{i-1}\right)}, ~ &\text{if}~\mu\in\mathcal{S}(u_i)\\
&0, ~&\text{if}~\mu\notin\mathcal{S}(u_i)
\end{aligned}\right.\nonumber \\
i=1,\dots,n_c, ~\mu\in\text{GF}(q)\backslash \hat{u}_i.
\end{align}
If $\hat{u}^{n_c}$ passes the CRC, the decoding is finished. In case the CRC fails, the SCF algorithm is given maximum $T_\text{max}$ attempts to correct the first symbol error. At the $T$-th attempt ($1\leq T\leq T_\text{max}$), the decoder finds the $T$-th largest element in $\boldsymbol{\varLambda}$. In case $\boldsymbol{\varLambda}_{i,\mu}$ is found, the SCF algorithm restarts the SC decoder by changing its estimate $\hat{u}_i$ to $\mu$. The CRC is checked after each attempt. This decoding process continues until the CRC passes or $T_\text{max}$ is reached. Note that the SCO-1 performance is a lower bound for SCF decoding. 

The performance and the average number of attempts ($T_\text{avg}$) of SCF on a biAWGN channel is shown in Fig.~\ref{fig:scf}. A $16$ bit CRC is used for error detection. We observe that SCF with $T_\text{max}=50$ performs very close to the SCO-1 bound. The average complexity and latency ($T_\text{avg}+1$) converge to SC in the high SNR regime ($\geq\SI{1.75}{dB}$), i.e., $T_\text{avg}+1\approx1$.
\begin{figure}
	\centering
\begin{tikzpicture}
\footnotesize
\begin{axis}[
ybar,
width=3in,
height=1.5in,
bar width=0.15cm,
symbolic x coords={1,2,3,4,5},
xtick=data,
grid=both,
xlabel = Number of errors caused by the channel,
ylabel = frequency of occurrence,
ymin=0,
ymax=1,
]
\addplot[fill=red!80!white,red!80!white] table[x=e,y=n]{e n
	1 0.91155
	2 0.08121
	3 0.00645
	4 7.2e-04
	5 7e-05
	};\addlegendentry{$E_s/N_0=\SI{1.75}{dB}$}
	
\addplot[fill=blue,blue] table[x=e,y=n]{e n
	1 0.95515
	2 0.04284
	3 0.00185
	4 1.6e-04
	5 0
};\addlegendentry{$E_s/N_0=\SI{2}{dB}$}

\addplot[fill=brown,brown] table[x=e,y=n]{e n
	1 0.98078
	2 0.01875
	3 4.6e-04
	4 1e-05
	5 0
};\addlegendentry{$E_s/N_0=\SI{2.25}{dB}$}

\end{axis}
\end{tikzpicture}
	\caption{Frequency of the number of errors caused by the channel for polar codes over GF($256$), $n_c=128$, $k=512$, $\text{primitive polynomial}=285$, $\alpha=29$, $\beta=1$, frozen position designed by Monte-Carlo method for SC decoding at $\SI{2.5}{dB}$}
	\label{fig:scf_errors}
\end{figure}
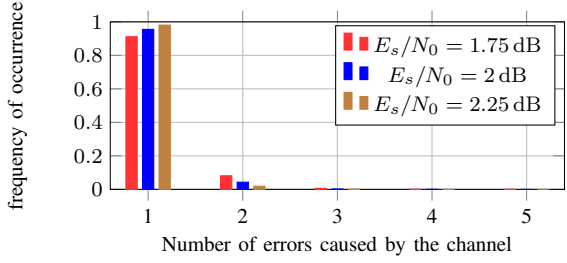
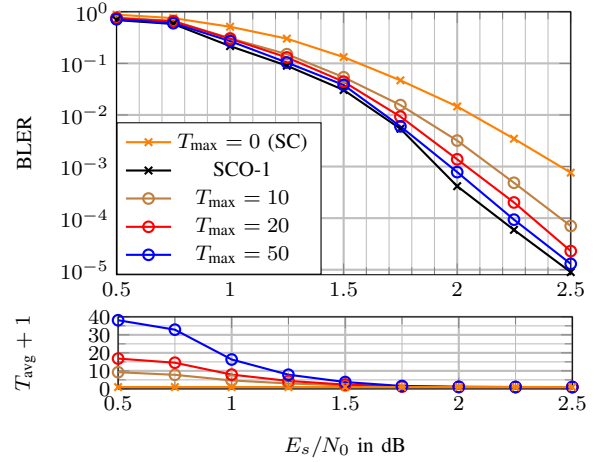
\begin{figure}
	\centering
	\begin{tikzpicture}[scale=1]
	\footnotesize
	\begin{semilogyaxis}[
	legend style={at={(0,0)},anchor=south west},
	ymin=0.000008,
	ymax=1,
	width=3in,
	height=2in,
	minor x tick num=4,
	minor y tick num=5,
	major grid style={thick},
	ytick={1,0.1,0.01,0.001,0.0001,0.00001,0.000001},
	grid=both,
	xmin = 0.5,
	xmax = 2.5,
	ylabel = BLER,
	]
	\addplot[orange, mark = x,thick]
	table[x=snr,y=fer]{snr fer
		0 1
		0.25 0.96875
		0.5 0.87931
		0.75 0.75325
		1 0.51282
		1.25 0.30157
		1.5 0.13234
		1.75 0.046871
		2 0.014612
		2.25 0.0034612
		2.5 0.0007612
	};\addlegendentry{$T_\text{max}=0$ (SC)}
	\addplot[black, mark = x,thick]
	table[x=snr,y=fer]{snr fer
		0 0.96774
		0.25 0.96774
		0.5 0.69048
		0.75 0.58
		1 0.21622
		1.25 0.089965
		1.5 0.030496
		1.75 0.0054076
		2 0.00041607
		2.25 5.9237e-05
		2.5 9e-06
	};\addlegendentry{SCO-1}
	\addplot[brown, mark = o,thick]
	table[x=snr,y=fer]{snr fer
		0 0.96774
		0.25 0.9375
		0.5 0.76923
		0.75 0.66667
		1 0.30612
		1.25 0.15075
		1.5 0.05386
		1.75 0.01556
		2 0.003183
		2.25 0.00048479
		2.5 7.0311e-05
	};\addlegendentry{$T_\text{max}=10$}
	
	\addplot[red, mark = o,thick]
	table[x=snr,y=fer]{snr fer
		0 0.96774
		0.25 0.9375
		0.5 0.75
		0.75 0.65217
		1 0.3
		1.25 0.13043
		1.5 0.044118
		1.75 0.0092764
		2 0.0013807
		2.25 0.00020121
		2.5 2.3e-05
	};\addlegendentry{$T_\text{max}=20$}
	
	\addplot[blue, mark = o,thick]
	table[x=snr,y=fer]{snr fer
		0 0.96774
		0.25 0.96774
		0.5 0.71429
		0.75 0.6
		1 0.27027
		1.25 0.10381
		1.5 0.038119
		1.75 0.0060084
		2 0.00078013
		2.25 9.3533e-05
		2.5 1.3e-05
	};\addlegendentry{$T_\text{max}=50$}	
	\end{semilogyaxis}
	\end{tikzpicture}
	\begin{tikzpicture}[scale=1]
	\footnotesize
	\begin{axis}[
	width=3in,
	height=1in,
	legend style={at={(1,1)},anchor=north east},
	legend columns=1, 
	ymin=0,
	ymax=40,
	minor x tick num=1,
	minor y tick num=1,
	major grid style={thick},
	grid=both,
	xmin = 0.5,
	xmax = 2.5,
	xlabel = $E_s/N_0\ \text{in dB}$,
	ylabel = $T_\text{avg}+1$,
	]
	\addplot[brown, mark = o,thick]
	table[x=snr,y=fer]{snr fer
		0 10.7097
		0.25 10.4688
		0.5 9.2564
		0.75 7.8
		1 4.6735
		1.25 3.0101
		1.5 1.9228
		1.75 1.2806
		2 1.0601
		2.25 1.0146
		2.5 1.0027
		
	};
	\addplot[red, mark = o,thick]
	table[x=snr,y=fer]{snr fer
		0 20.387
		0.25 19.844
		0.5 16.825
		0.75 14.522
		1 7.92
		1.25 4.4087
		1.5 2.3926
		1.75 1.3516
		2 1.0789
		2.25 1.0174
		2.5 1.0032
	};
	
	\addplot[blue, mark = o,thick]
	table[x=snr,y=fer]{snr fer
		0 49.419
		0.25 49.387
		0.5 38.119
		0.75 32.88
		1 16.315
		1.25 7.8651
		1.5 3.7929
		1.75 1.5946
		2 1.1066
		2.25 1.0216
		2.5 1.0038
	};
	
	\addplot[orange, mark = x,thick]
	table[x=snr,y=fer]{snr fer
		0 1
		0.25 1
		0.5 1
		0.75 1
		1 1
		1.25 1
		1.5 1
		1.75 1
		2 1
		2.25 1
		2.5 1
	};
	\end{axis}
	\end{tikzpicture}
	\caption{SCF decoding performance of polar codes over GF($256$), $n_c=128$, $k=512$, $\ell_\text{CRC}=16$, $\text{primitive polynomial}=285$, $\alpha=29$, $\beta=1$, frozen position designed by Monte-Carlo method for SC decoding at $\SI{2.5}{dB}$}
	\label{fig:scf}
\end{figure}

\subsection{SCL Decoding and SCL with a Pruned Tree}\label{subsec:prun}
SCL decoding was proposed in~\cite{tal2015list} and improves the performance of SC decoding by deploying $L$ parallel SC decoding paths. The reliability of each path is described by a path metric (PM). For $q$-ary polar codes, the PM of $\hat{u}^i$ and the recursive update rule is given by
\begin{align}
\text{PM}(\hat{u}^i)&=P_{U^i|Y^n}\left(\hat{u}^i|y^n\right)\nonumber \\
&=P_{U_i|Y^nU^{i-1}}\left(\hat{u}_i|y^n\hat{u}^{i-1}\right)P_{U^{i-1}|Y^n}\left(\hat{u}^{i-1}|y^n\right)\\
&=\underbrace{P_{U_i|Y^nU^{i-1}}\left(\hat{u}_i|y^n\hat{u}^{i-1}\right)}_{\text{from (\ref{eq:sc})}}\text{PM}(\hat{u}^{i-1}) \label{eq:pm}
\end{align}
where $\text{PM}(\hat{u}^0)=\text{PM}(\varnothing)$ is initialized to $1$. Assuming $L_{i-1}$ paths of length $i-1$ survive at the begin of decoding phase $i$. The SCL decoder first uses $L_{i-1}$ parallel SC decoders to compute the PMF in (\ref{eq:sc}) for all survived paths. Then $|\mathcal{S}(u_i)|L_{i-1}$ PMs are computed by (\ref{eq:pm}). The most likely $L_i$ paths are selected as survivors and passed into phase $i+1$, where 
$
L_i=\min\left(L,|\mathcal{S}(u_i)|L_{i-1}\right).
$ In our implementation, the \enquote{lazy copy} strategy is deployed to reduce copy operations and the PMs are normalized in the same manner as in~\cite[Algorithm 10, lines 20-25]{tal2015list} to avoid numerical problems.

The complexity of SCL decoding is dominated by the maximum list size $L$. In this work, we use the number of visited nodes in the decoding tree ($N_\text{node}$) to evaluate the SCL decoding complexity. For conventional SCL decoding, we have $N_\text{node}=\sum_{i=1}^{n_c}L_i.$ We observe that $N_\text{node}$ is independent of the channel quality, which means the conventional SCL decoder does not adapt its complexity for different SNRs.

We propose an approach to adapt the complexity by pruning the decoding tree of the conventional SCL algorithm. The basic idea is to delete \enquote{unreliable} decoding paths although they are the $L_i$ most likely. Let $\text{SPM}_i\left[l\right],~l=1,\dots,L_i$ denote the sorted $L_i$ largest PMs ($\text{SPM}_i\left[1\right]$ is the largest). The path indexed by $l$ is eliminated if at least one of the following conditions is fulfilled:
\begin{itemize}
	\item[1.] $\text{SPM}_i\left[l\right] < \delta_1 \text{SPM}_i\left[1\right]$
	\item[2.] $\text{SPM}_i\left[l\right] < \delta_2 \text{SPM}_i\left[l-1\right]$ 
\end{itemize}
where $0\leq\delta_1\leq\delta_2\leq1.$
The predefined parameters $\delta_1$ and $\delta_2$ describe a reliability requirement. If $\delta_1=\delta_2=0$, this approach is equivalent to the conventional SCL decoder. If $\delta_1=\delta_2=1$, this approach is equivalent to the SC decoder. 

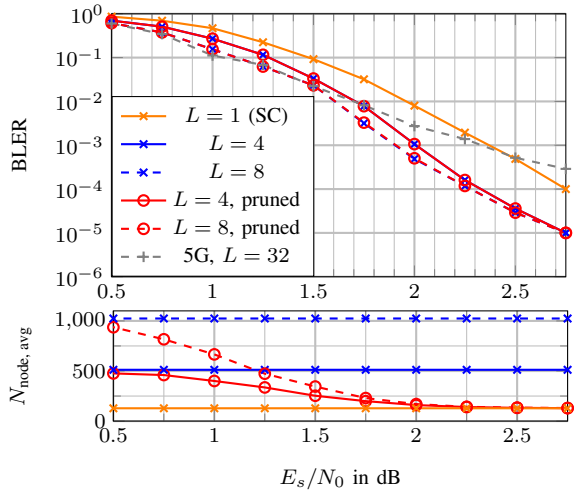
\begin{figure}
	\centering
	\begin{tikzpicture}[scale=1]
	\footnotesize
	\begin{semilogyaxis}[
	legend style={at={(0,0)},anchor=south west},
	ymin=0.000001,
	ymax=1,
	width=3in,
	height=2in,
	minor x tick num=4,
	minor y tick num=5,
	major grid style={thick},
	ytick={1,0.1,0.01,0.001,0.0001,0.00001,0.000001},
	grid=both,
	xmin = 0.5,
	xmax = 2.75,
	ylabel = BLER,
	]
		\addplot[orange, mark = x,thick]
		table[x=snr,y=fer]{snr fer
			0 1
			0.25 0.97059
			0.5 0.86047
			0.75 0.69492
			1 0.46847
			1.25 0.22394
			1.5 0.09212
			1.75 0.031973
			2 0.0080037
			2.25 0.001941
			2.5 0.000488
			2.75 0.0001
		};\addlegendentry{$L=1$ (SC)}
	\addplot[blue, mark = x,thick]
	table[x=snr,y=fer]{snr fer
		0 1
		0.25 0.88235
		0.5 0.69767
		0.75 0.50847
		1 0.27027
		1.25 0.11583
		1.5 0.033296
		1.75 0.0077983
		2 0.0010624
		2.25 0.00016086
		2.5 3.6e-05
		2.75 1e-05
	};\addlegendentry{$L=4$}
	\addplot[blue, mark = x, mark options={solid}, dashed,thick]
	table[x=snr,y=fer]{snr fer
		0 0.96774
		0.25 0.90909
		0.5 0.61224
		0.75 0.375
		1 0.15464
		1.25 0.0625
		1.5 0.023274
		1.75 0.003175
		2 0.00048718
		2.25 0.00011789
		2.5 2.9e-05
		2.75 1e-05
	};\addlegendentry{$L=8$}
	\addplot[red, mark = o,thick]
	table[x=snr,y=fer]{snr fer
		0 1
		0.25 0.88235
		0.5 0.69767
		0.75 0.50847
		1 0.27027
		1.25 0.11583
		1.5 0.033296
		1.75 0.0077983
		2 0.0010624
		2.25 0.00016086
		2.5 3.6e-05
		2.75 1e-05
	};\addlegendentry{$L=4$, pruned}
	\addplot[red, mark = o, mark options={solid}, dashed,thick]
	table[x=snr,y=fer]{snr fer
		0 0.96774
		0.25 0.90909
		0.5 0.61224
		0.75 0.375
		1 0.15464
		1.25 0.0625
		1.5 0.023274
		1.75 0.0032844
		2 0.00050398
		2.25 0.00011789
		2.5 2.9e-05
		2.75 1e-05
	};\addlegendentry{$L=8$, pruned}
	\addplot[mark options={solid}, mark = +,gray,dashed,thick]
	table[x=snr,y=fer]{snr fer
		0.5 0.6
		0.75 0.34483
		1 0.11152
		1.25 0.068027
		1.5 0.022305
		1.75 0.0081699
		2 0.002743
		2.25 0.0013824
		2.5 0.0005183
		2.75 0.00028897
	};\addlegendentry{5G, $L=32$}
	\end{semilogyaxis}
	\end{tikzpicture}
	\begin{tikzpicture}[scale=1]
	\footnotesize
	\begin{axis}[
	width=3in,
	height=1.2in,
	legend style={at={(1,1)},anchor=north east},
	legend columns=1, 
	ymin=0,
	ymax=1100,
	minor x tick num=1,
	minor y tick num=1,
	major grid style={thick},
	grid=both,
	xmin = 0.5,
	xmax = 2.75,
	xlabel = $E_s/N_0\ \text{in dB}$,
	ylabel = $N_\text{node, avg}$,
	]
	\addplot[red, mark = o,thick]
	table[x=snr,y=fer]{snr fer
		0 512
		0.25 504.4800
		0.5 475.9
		0.75 459.66
		1 400.36
		1.25 336
		1.5 253
		1.75 199.2140
		2 159.970
		2.25 140.4840
		2.5 132.1680
		2.75 129.290
	};
	
	\addplot[blue, mark = x,thick]
	table[x=snr,y=fer]{snr fer
		0 512
		0.25 512
		0.5 512
		0.75 512
		1 512
		1.25 512
		1.5 512
		1.75 512
		2 512
		2.25 512
		2.5 512
		2.75 512
	};
	
	\addplot[red, mark = o, mark options={solid}, dashed,thick]
	table[x=snr,y=fer]{snr fer
		0 1021.7
		0.25 1017.5
		0.5 936.70
		0.75 816.42
		1 665.74
		1.25 473.12
		1.5 344.74
		1.75 229.880
		2 169.752
		2.25 142.426
		2.5 134.468
		2.75 130.29
	};
	
	\addplot[blue, mark = x, mark options={solid}, dashed,thick]
	table[x=snr,y=fer]{snr fer
		0 1024
		0.25 1024
		0.5 1024
		0.75 1024
		1 1024
		1.25 1024
		1.5 1024
		1.75 1024
		2 1024
		2.25 1024
		2.5 1024
		2.75 1024
	};
	
	\addplot[orange, mark = x,thick]
	table[x=snr,y=fer]{snr fer
		0 128
		0.25 128
		0.5 128
		0.75 128
		1 128
		1.25 128
		1.5 128
		1.75 128
		2 128
		2.25 128
		2.5 128
		2.75 128
	};
	
	\end{axis}
	\end{tikzpicture}
	\caption{SCL decoding and SCL with a pruned tree for polar codes over GF($256$), $n_c=128$, $k=512$, no CRC, $\text{primitive polynomial}=285$, $\alpha=29$, $\beta=1$, $\delta_1=10^{-6}$, $\delta_2=10^{-5}$, frozen position designed by Monte-Carlo method for SC decoding at $\SI{2.5}{dB}$}
	\label{fig:scl}
\end{figure}

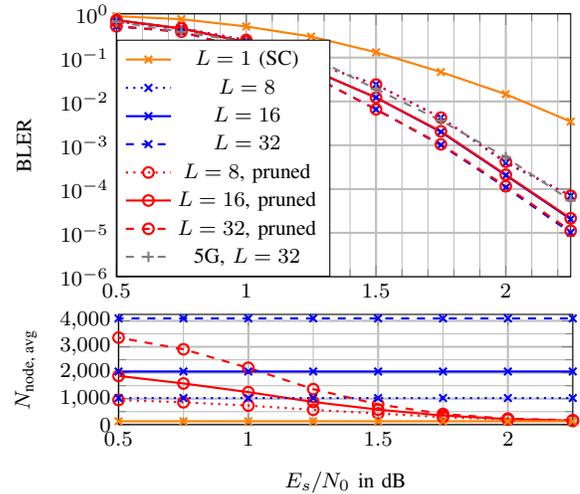
\begin{figure}
	\centering
	\begin{tikzpicture}[scale=1]
	\footnotesize
	\begin{semilogyaxis}[
	legend style={at={(0,0)},anchor=south west},
	ymin=0.000001,
	ymax=1,
	width=3in,
	height=2in,
	minor x tick num=4,
	minor y tick num=5,
	major grid style={thick},
	ytick={1,0.1,0.01,0.001,0.0001,0.00001,0.000001},
	grid=both,
	xmin = 0.5,
	xmax = 2.25,
	ylabel = BLER,
	]
	\addplot[orange, mark = x,thick]
	table[x=snr,y=fer]{snr fer
		0 1
		0.25 0.96875
		0.5 0.87931
		0.75 0.75325
		1 0.51282
		1.25 0.30157
		1.5 0.13234
		1.75 0.046871
		2 0.014612
		2.25 0.0034612
	};\addlegendentry{$L=1$ (SC)}
	\addplot[mark options={solid},dotted,blue, mark = x,thick]
	table[x=snr,y=fer]{snr fer
		0 0.96774
		0.25 0.9375
		0.5 0.66667
		0.75 0.46875
		1 0.2521
		1.25 0.077922
		1.5 0.024272
		1.75 0.0042638
		2 0.00039561
		2.25 7.023e-05
	};\addlegendentry{$L=8$}
	\addplot[blue, mark = x,thick]
	table[x=snr,y=fer]{snr fer
		0 1
		0.25 0.85714
		0.5 0.71429
		0.75 0.46154
		1 0.22556
		1.25 0.053381
		1.5 0.012255
		1.75 0.0020084
		2 0.00020555
		2.25 0.00002055
	};\addlegendentry{$L=16$}
	\addplot[mark options={solid},dashed,blue, mark = x,thick]
	table[x=snr,y=fer]{snr fer
		0 0.96774
		0.25 0.9375
		0.5 0.51724
		0.75 0.38961
		1 0.15385
		1.25 0.039164
		1.5 0.0065949
		1.75 0.0010121
		2 0.00010829
		2.25 0.0000102
	};\addlegendentry{$L=32$}
	\addplot[mark options={solid},dotted,red, mark = o, thick]
	table[x=snr,y=fer]{snr fer
		0 0.96774
		0.25 0.9375
		0.5 0.66667
		0.75 0.46875
		1 0.2521
		1.25 0.077922
		1.5 0.024272
		1.75 0.0042638
		2 0.00042387
		2.25 7.023e-05
	};\addlegendentry{$L=8$, pruned}
	\addplot[red, mark = o,thick]
	table[x=snr,y=fer]{snr fer
		0 1
		0.25 0.85714
		0.5 0.71429
		0.75 0.46154
		1 0.22556
		1.25 0.053381
		1.5 0.012255
		1.75 0.0020777
		2 0.00021264
		2.25 0.00002155
	};\addlegendentry{$L=16$, pruned}
	\addplot[mark options={solid},dashed,red, mark = o,thick]
	table[x=snr,y=fer]{snr fer
		0 0.96774
		0.25 0.9375
		0.5 0.51724
		0.75 0.38961
		1 0.15385
		1.25 0.039164
		1.5 0.0065949
		1.75 0.001047
		2 0.00011603
		2.25 0.0000112
	};\addlegendentry{$L=32$, pruned}
		\addplot[mark options={solid}, mark = +,gray,dashed,thick]
		table[x=snr,y=fer]{snr fer
			0.5 0.65217
			0.75 0.38961
			1 0.22388
			1.25 0.10417
			1.5 0.01906
			1.75 0.0036492
			2 0.00053108
			2.25 5.5e-05
		};\addlegendentry{5G, $L=32$}
	\end{semilogyaxis}
	\end{tikzpicture}
	\begin{tikzpicture}[scale=1]
	\footnotesize
	\begin{axis}[
	width=3in,
	height=1.2in,
	legend style={at={(1,1)},anchor=north east},
	legend columns=1, 
	ymin=0,
	ymax=4250,
	minor x tick num=1,
	minor y tick num=1,
	major grid style={thick},
	grid=both,
	xmin = 0.5,
	xmax = 2.25,
	xlabel = $E_s/N_0\ \text{in dB}$,
	ylabel = $N_\text{node, avg}$,
	]
		\addplot[mark options={solid},dotted,red, mark = o,thick]
		table[x=snr,y=fer]{snr fer
			0 1024
			0.25 1014.1
			0.5 948.79
			0.75 860.82
			1 733.47
			1.25 571.11
			1.5 429.05
			1.75 286.63
			2 198.54
			2.25 144.36
		};
	\addplot[red, mark = o,thick]
	table[x=snr,y=fer]{snr fer
		0 2048
		0.25 1967.3
		0.5 1877.1
		0.75 1586.5
		1 1252.3
		1.25 873.05
		1.5 576.9
		1.75 348.63
		2 214.01
		2.25 149.06
	};
	\addplot[mark options={solid},dashed,red, mark = o,thick]
	table[x=snr,y=fer]{snr fer
		0 4093.3
		0.25 4027.9
		0.5 3353.6
		0.75 2905
		1 2188.4
		1.25 1367.6
		1.5 779.8497
		1.75 409.9297
		2 227.8594
		2.25 169.0642
	};
	\addplot[blue, mark = x,thick]
	table[x=snr,y=fer]{snr fer
		0 2048
		0.25 2048
		0.5 2048
		0.75 2048
		1 2048
		1.25 2048
		1.5 2048
		1.75 2048
		2 2048
		2.25 2048
	};
	\addplot[mark options={solid},dotted,blue, mark = x,thick]
	table[x=snr,y=fer]{snr fer
		0 1024
		0.25 1024
		0.5 1024
		0.75 1024
		1 1024
		1.25 1024
		1.5 1024
		1.75 1024
		2 1024
		2.25 1024
	};
	\addplot[mark options={solid},dashed,blue, mark = x,thick]
	table[x=snr,y=fer]{snr fer
		0 4096
		0.25 4096
		0.5 4096
		0.75 4096
		1 4096
		1.25 4096
		1.5 4096
		1.75 4096
		2 4096
		2.25 4096
	};
	
	\addplot[orange, mark = x,thick]
	table[x=snr,y=fer]{snr fer
		0 128
		0.25 128
		0.5 128
		0.75 128
		1 128
		1.25 128
		1.5 128
		1.75 128
		2 128
		2.25 128
	};
	
	\end{axis}
	\end{tikzpicture}
	\caption{SCL decoding and SCL with a pruned tree for polar codes over GF($256$), $n_c=128$, $k=512$, $\ell_\text{CRC}=16$, $\text{primitive polynomial}=285$, $\alpha=29$, $\beta=1$, $\delta_1=10^{-6}$, $\delta_2=10^{-5}$, frozen position designed by Monte-Carlo method for SC decoding at $\SI{2.5}{dB}$}
	\label{fig:sclcrc}
\end{figure}

The performance and decoding complexity of SCL are shown in Fig.~\ref{fig:scl} (without CRC) and Fig.~\ref{fig:sclcrc} (with CRC). We observe that the proposed SCL with a pruned decoding tree performs very close to the conventional SCL with significantly lower complexity. In the high SNR regime ($\geq\SI{2.25}{dB}$), the average complexity converges to SC, i.e., $N_\text{node, avg}\approx N_\text{node,SC} =n_c$. We also provide simulation results for binary polar codes in 5G with the same code length $n_{c,\text{bin}}$, dimension $k$ and outer codes. Fig.~\ref{fig:scl} shows that polar codes over GF(256) provide a better SC performance and distance properties than binary polar codes. In Fig.~\ref{fig:sclcrc}, we observe that GF(256) polar codes can achieve the same performance of binary polar codes with the list size reduced by a factor of $4$.

\section{Conclusion and Future Work}\label{sec:con}
We discussed non-binary polar codes based on $q$-ary $2\times2$ kernels. The kernel is selected by maximizing the \enquote{single-level} polarization effect. The performance and complexity of BP, SCF and SCL decoding algorithms based on $q$-ary message passing are analyzed. We proposed an SCL decoder with a pruned tree to adapt the decoding complexity. 

A non-Monte-Carlo based code design (selection of frozen positions) and the analysis of non-binary polar codes for higher-order transmission will be investigated in the future. 

\bibliographystyle{IEEEtran}
\bibliography{IEEEabrv,references}
\end{document}